\documentclass[aps,prl,preprint,groupedaddress]{revtex4}

\usepackage{graphicx,amsmath,amssymb}

\newtheorem{Def}{Definition}[section]

\begin{document}

\title{Separable local fractional differential equations }

\author{Kiran M. Kolwankar}
\email[]{Kiran.Kolwankar@gmail.com}
\address{Department of Physics, Ramniranjan Jhunjhunwala College, Ghatkopar(W) Mumbai 400086}

\date{\today}

\begin{abstract}
The concept of local fractional derivative was introduced in order to be able
to study the local scaling behavior of functions. However it has turned out to be much more
useful. It was found that simple equations involving these operators naturally  
incorporate the fractal sets into the equations. Here, the scope of these equations
has been extended further by considering different possibilities for the known function.
We have also studied a separable local fractional differential equation along with
its method of solution.
\end{abstract}



\maketitle

\section{Introduction and definitions}

The derivatives and integrals of non-integer 
order~\cite{2OS,POD,SAM} 
have been found to be useful in successfully describing scaling processes.
The realm of applications of such a fractional calculus is fast
expanding with ever new developments rapidly
taking place in the field of statistical and nonlinear physics
over last few years~\cite{MK1,Hil,MK2,Zas,
WBG,GG,CM}.
The most often used definition
of the fractional derivative is through the
so called
Riemann-Liouville fractional derivative~\cite{2OS,SAM,POD}. 
For $q$, the order of the derivative, between zero
and one it is given by:
\begin{Def} The Riemann-Liouville fractional derivative of a function $f$ of order $q$ ($0<q<1$) is defined
as:
\begin{eqnarray}
D^q_xf(x') &=& \left\{\begin{array}{ll}
D^q_{x+}f(x'), & x'>x, \\
D^q_{x-}f(x'), & x' < x. 
 \end{array}\right. \nonumber \\
&=& \frac{1}{\Gamma(1-q)}
\left\{\begin{array}{ll}
\frac{d}{dx'}\int_x^{x'}f(t)(x'-t)^{-q} dt, & x'>x, \\
-\frac{d}{dx'}\int_{x'}^xf(t)(t-x')^{-q} dt, & x' < x. 
 \end{array}\right.
\end{eqnarray}

\end{Def}

Clearly this is nonlocal and depends on the limit $x$. 
The value of $x$ is usually dictated by the 
problem one is investigating.
In some cases it is appropriate to put it equal to 
$\pm \infty$. In which case it is called Weyl derivative. Another
choice of the limit is made in the following definition called the
Local fractional derivative (LFD) defined as 
follows~\cite{2KG1}:
\begin{Def} The local fractional derivative of order $q$ ($0<q<1$) of a function $f\in C^0: \mathbb{R} \rightarrow \mathbb{R}$
is defined as \[ {\cal{D}}^q f(x) = \lim_{x'\rightarrow x} D^q_x(f(x')-f(x)) \] if the limit exists in $\mathbb{R}$.

\end{Def}
Such a local definition of fractional derivative is useful in studying the local
scaling behavior in fractal and multifractal functions.
This definition naturally appears in the local fractional Taylor
expansion~\cite{2KG1} 
giving it a geometrical interpretation which is a simple extension of that of ordinary derivatives.
That is, just as the first order derivative is a coefficient of the local linear approximation, the LFD
 is the coefficient of the local power law approximation. It should be noted
that the extra limit in the definition of the LFD makes it very different
from other definitions of fractional derivatives and some of its properties
are very different from the non-local versions of fractional derivatives
(see~\cite{AC,BDG,
CCK} for more mathematical properties).
One important
difference is that though it reduces to the usual derivative when $q=1$,
the LFD is not an analytic function of $q$ for a given
function. For example, if the function is smooth then
the LFD of any order $q$ less than one is zero. In fact, in general
for any continuous nondifferentiable function
there exists a critical order of differentiability 
between zero and one below which
all the derivatives are zero and above which they do not exist.
This also means that with this definition one can not take an LFD of
order greater than the critical order. In~\cite{Kol2} we have introduced a recursive
definition of LFD which allows us to overcome this problem. Also, as of now, we do not know
any example of a continuous everywhere but 
nowhere differentiable function for which the LFD 
exists at the critical
order. But here we take a point of view that such functions nevertheless
exist and hope that this aspect will receive rigorous treatment in future, possibly by
using nonstandard analysis. 
This critical order of differentiability is equivalent to
the local H\"older exponent
or the local power law exponent~\cite{KLV},
and for
this formalism to yield meaningful results we should work at the
order which is equal to this exponent. Several 
authors~\cite{KTAR,LDE,CC,CYZ,Wu,Yan,LZ} have
attempted to take this work further

\section{A simple local fractional differential equations}
Once we have the definition of  LFD, the next natural step is to consider equations involving LFD, the local fractional differential equation (LFDE). The simplest
such equation is
\begin{eqnarray}\label{eq:lfde}
{\cal{D}}^q f(x) = g(x)
\end{eqnarray}
where $g(x)$ is a known function and $f(x)$ an unknown. Using
the local fractional Taylor expansion~\cite{2KG1} 
its solution can be written as 
a generalised Riemann sum giving
\begin{eqnarray}\label{eq:soln}
f(x) = \int g(x) d^{q}x = \lim_{N\rightarrow\infty} \sum_{i=0}^{N-1} g(x_i^*)
{(x_{i+1}-x_i)^q\over\Gamma(q+1)}
\end{eqnarray}
where $g(x_i^*)$ is an appropriately chosen point in the interval
$[x_i,x_{i+1}]$. 
In~\cite{Kol1}, it was shown that this integral does not exist if $g(x)$ is a 
continuous function in any interval.
 
 Yet,  
two classes of functions $g(x)$ can be identified which will
yield a nontrivial function $f(x)$ as a solution. The first class corresponds
to the functions which have fractal support. 
An example $g(x)=1_C(x)$, where $1_C(x)$ is an indicator or characterstic
function of a Cantor set $C$, was considered in~\cite{2KG5}.
Then, if $q=\alpha$, the
dimension of the support of the function, the sum above converges 
since only a few terms contribute to the sum giving
rise to a finite solution. This function, called the "devil's
staircase", changes only on
the points of the Cantor set $C$ and is constant everywhere else.
We denote this solution by $P_C(x)$. 
The second class
consists of rapidly oscillating
functions which oscillate around zero in any small interval. 
These oscillations then result in cancellations  in the sum in~(\ref{eq:soln}) again giving
rise to a finite solution.
A realization of the white noise
is one example in this class of functions. This example has been
discussed in~\cite{KK1,Kol1}.

In this work, we are going to restrict to the first class discussed above, that
is, the one involving functions having fractal support. The treatment so far
is far from being rigorous and we have relied only on the handwaving arguments.
There are several issues involved here. It is easy to convince that the sum 
in~(\ref{eq:soln}) will lead to a devil's staircase like functions when $g(x)$ has
a fractal support but, for a rigorous
treatment, the limit has to be established by considering upper and lower
generalized Riemann sums. Furthermore, proving that the inverse also
holds is even more difficult. It is clear that the staircase like solution will
have local scaling (H\"older) exponent equal to $\alpha$, the dimension, 
at the points of the support. So the critical order of the solution
will be $\alpha$ which is consistent with the order of the original differential
equation. But as we have discussed before, it is difficult to show that the
LFD at the critical order exists. This difficulty arises owing to the oscillations
around the local power law which exist in fractal functions. But this is not
a serious point as these oscillations, as shown in~\cite{KK},  are small and do not affect the power
law exponent. But it becomes a hurdle in generalizing the
fundamental theorem of calculus to LFDs. 
A lot of efforts would be needed, possibly with minor modification of the
definition of LFD or a treatment using nonstandard analysis, to develope
the theory rigorously. In fact, Mandelbrot~\cite{1Man} has already advocated 
studying such integrals using nonstandard analysis.
In this work, we are not
going to get into these technical difficulties but instead develop the theory,
albeit symbolically, assuming that the quantities defined exist.

Carpinteri and Cornetti~\cite{CC} used the concept of LFD to introduce 
a relation between the strain and the displacement when the 
strain is localized on the fractal set. They proposed
\begin{eqnarray}\nonumber
\epsilon^*(x) = {\cal{D}}^\alpha u(x)
\end{eqnarray}
where $\epsilon^*(x)$ is a renormalised strain and $u(x)$ is the displacement.
This also offers a physical interpretation to LFD. This formalism is useful
in studying the structural properties of concrete-like or disordered materials.
In~\cite{CCK}, we calculated resultant of a stress distribution and its moment
when the stress is distributed over a fractal set. Further use of LFD and the fractal
integral for the principle of virtual work and to generalize the constitutive 
equations of elasticity has been made in~\cite{CCC, CCC1}.

\section{Separable local fractional differential equations}
In this section we develop the theory of
LFDEs further. First we consider more general forms of the
function $g(x)$ and then we treat more general right hand side in~(\ref{eq:lfde}) which
can depend even on the unknown function.

Now consider $g(x)=\bar{g}(x) 1_C(x)$ where $\bar{g}(x)$ is some sufficiently well-behaved
function. The multiplication by $1_C(x)$ makes $g(x)$ a function with fractal support.
In such a situation the solution $f(x)$ in Eq.~(\ref{eq:soln}) can in
principle be computed. The exact result will also depend on the fractal set supporting the
function. Some explicit calculations have been performed in~\cite{CCK} when $\bar{g}(x) = x$ or $x^2$ and
when the support is either a middle $1/3$ Cantor set or a middle $1/p$ Cantor set.
A similar method can be used for any power $x^n$. And if $\bar{g}(x)$ is real analytic then
one can use its power series representation and an approximating scheme for the solution~(\ref{eq:soln})
can be set up which can at least be evaluated numerically.

Next nontrivial step is to consider $g(x) = g_1(x) 1_{C_1}(x) + g_2(x) 1_{C_2}(x)$ where $C_1$
and $C_2$ are different fractal sets with the same dimension $\alpha$ and $g_1$, $g_2$ are
different functions. Clearly, the integral
operator in~(\ref{eq:soln}) is a linear operator and the result of considering such a $g(x)$
in the solution is an addition of two devil's staircase like functions coming from two individual
terms in above expression for $g(x)$. It is important to realise that this is an interesting 
new step since it allows to incorporate effect
of two different processes on two different fractal sets in one single equation and is not just another
devil's staircase on a simple union of two sets. That is, the 
$g(x)$ can not be written as $(g_1(x) + g_2(x) )1_{C_1 \cup C_2}(x)$ even when $C_1$
and $C_2$ are disjoint.
It is not difficult to imagine physical situations where
such a function is needed. For example, we can generalise the situation discussed in~\cite{2KG5}
in which a local fractional diffusion equation was solved for a diffusion process taking place
in fractal time. Now suppose that there are two different processes happening simultaneously
but on two different fractal time sets of the same dimension but possibly with different lacunarity and with different constants of diffusion. This could be a result of two different kind of traps. This 
choice of the known function will allow us to incorporate such situations into equations.

Now, in order to incorporate further complex situations, we rewrite the Eq.~(\ref{eq:lfde}) using a different notation and also generalize it.
We write
\begin{eqnarray}\label{eq:glfde}
\frac{dy}{dx^q} = g(x,y).
\end{eqnarray}
Here the LHS is a mere change in notation where LFD is written in more suggestive form. 
First of all $y$ is now our unknown function to be found out. The order of the LFD
$q$ appears only in the denominator since the numerator is only a first order difference.
The RHS is now a generalization wherein the known function $g$ depends also on $y$.
Now we assume that the equation is separable, that is, $g(x,y)= \bar{g}(x) 1_C(x) h(y)$
a product of functions which depend only on $x$ or $y$. This also means that
the fractal set $C$ does not change with $y$. Therefore we have
\begin{eqnarray}\label{eq:slfde}
\frac{dy}{dx^q} = \bar{g}(x) 1_C(x) h(y).
\end{eqnarray}
We can formally rewrite this equations as
\begin{eqnarray}\label{eq:solslfde}
\frac{dy}{h(y)} = \bar{g}(x) 1_C(x)dx^q
\end{eqnarray}
and integrate to obtain the solution apart from the constant of integration
\begin{eqnarray}\label{eq:intslfde}
\int \frac{dy}{h(y)} = \int \bar{g}(x) 1_C(x)dx^q
\end{eqnarray}
where the RHS is now the integral as in Eq.~(\ref{eq:soln}).
These steps can be justified using the chain rule for LFD, proved in~\cite{AC}, 
as in the following. Owing to the chain rule we can write 
two auxilary equations in place of Eq.~(\ref{eq:slfde}):
\begin{eqnarray}
\frac{dy}{du} = h(y) \;\;\; \mbox{and} \;\;\; \frac{du}{dx^q} = \bar{g}(x) 1_C(x).
\end{eqnarray}
These equations can be solved separately and then the solutions can be
combined to obtain the solution~(\ref{eq:intslfde}).

As an example, consider $h(y)=y$ and $\bar{g}(x) = 1$. In this case,
the solution becomes $y= A \exp(P_C(x))$. A similar case had arisen
in the solution of the local fractional Fokker-Planck equation in~\cite{2KG5}
though, over there, it was solved using different steps.
As another example, if we consider $h(y) = y^2$ keeping the $\bar{g}(x)=1$ then
the solution is $y = - 1/{P_C(x)}$. This demonstrates that complex situations 
involving fractals can be incorporated in a differential equation and solutions
can be obtained which includes the information of the fractal set in the form
of the function $P_C(x)$.

\section{Concluding discussions}
In this paper we have briefly reviewed the introduction to LFD and
related developments regarding the equations involving the LFDs. The 
simplest example of the so called LFDE was also recalled. The main purpose
of this paper was to take these developments further. To this end, we 
have step by step increased the complexity in the RHS of the simple LFDE.
From the indicator function of the Cantor set we first considered the case
of a function multiplied by the indicator function and discussed its solution.
Then we took a next nontrivial step consisting of addition of two functions
with different fractal sets of the same dimension as the support. Lastly,
we studied a more general function which depended on both $x$ and $y$
but was separable, that is, could be written as a product of functions only
of $x$ and only of $y$. We discussed and rationalized the method of its
solution. The examples considered demonstrate the utility of this approach
in studying the process taking place on a fractal set. One can also add a
white noise to the known function $g$ on the RHS of Eqs.~(\ref{eq:lfde})
or~(\ref{eq:glfde}) which will yield a random process with additional 
log-periodic oscillations. Such stochastic differential equations will be
considered elsewhere.

Taking this line of exploration further will be interesting and also challenging.
Also, it is necessary to establish these results rigorously. These developments
will pave a way for fractal dynamical systems.

\section{Acknowledgement}
The author would like to thank Council for Scientific and Industrial Research, India
for financial assistance.


\bibliographystyle{plain}

\end{document}